# Is the epidemic spread related to GDP? Visualizing the distribution of COVID-19 in Chinese Mainland

**Yi ZHANG**

College of Geography and Environmental Science, Northwest Normal University, China; Institute of Earth Environment, Chinese Academy of Sciences, China

**Hanwen TIAN**

College of Geography and Environmental Science, Northwest Normal University, China

**Yinglong ZHANG**

Zhejiang Provincial Institute of Communications Planning, Design & Research Co., Ltd., China

**Yiping CHEN**

Institute of Earth Environment, Chinese Academy of Sciences, China

## Abstract

In December 2019, COVID-19 were detected in Wuhan City, Hubei Province of China. SARS-CoV-2 rapidly spread to the whole Chinese mainland with the people during the Chinese Spring Festival Travel Rush. As of 19 February 2020, 74576 confirmed cases of COVID-19 had been reported in Chinese Mainland. What kind of cities have more confirmed cases, and is there any relationship between GDP and confirmed cases? In this study, we explored the relationship between the confirmed cases of COVID-19 and GDP at the prefectural-level, found a positive correlation between them. This finding warns high GDP areas should pay more prevention and control efforts when an epidemic outbreak, as they have greater risks than other areas nearby.

**Epidemic globally spread in 21st century**

Humans are attacked by various epidemic diseases, causing great disaster to human beings. Only in the 21st century, SARS, MERS and other diseases have spread globally, and a large number of people have died. According to World Health Organization (WHO), the SARS virus in 2003 caused a total of 8,096 SARS cases worldwide, killed 774 people, involved 29 countries and regions (WHO, 2004). And the MERS virus caused a total of 2494 cases worldwide, killed 858 people since April 2012 (WHO, 2019).

  At the end of 2019, several unexplained pneumonia cases were found in Wuhan, Hubei Province, China. The clinical manifestations were mainly fever, combined with symptoms such as dry cough, fatigue, poor breathing, diarrhea, runny nose, and sputum. Some patients experienced dyspnea after one week, a large amount of fibrous exudation in the lungs, and pulmonary fibrosis lesions. In severe cases, they even died of shock (Huang et al., 2020). In January 2020, researches revealed that a new type of coronavirus caused the pneumonia (WHO, 2020a). In February 2020, The International Committee on Taxonomy of Viruses (ICTV) officially named the virus SARS-CoV-2, and WHO named pneumonia caused by the virus as COVID-19.

**COVID-19 cases in China**

The time of COVID-19 outbreak is at the Chinese Spring Festival Travel Rush (Chunyun). During this time, many people back their hometown for celebrating Spring Festival, and the virus quickly spread from Wuhan to other parts of China with the flow of people. On January 29, 31 provinces in the Chinese mainland launched the first level response mechanism for public health events (the highest). In order to prevent the further spread of the epidemic, the Chinese government decided to begin strict controls on entering and leaving Wuhan from January 23, and since then personnel have rarely continued to flow from Wuhan. As of two incubation cycles since Wuhan was sealed on January 23, that is, February 19, a total of 74,577 cases of COVID-19 were confirmed in the Chinese mainland (National Health Commission of the People's Republic of China, 2020). No matter in Hubei Province where the outbreak originated or in other provinces, the epidemic situation varies greatly in their respective regions.

As an important indicator of economic, GDP can also be used to feedback the vitality of various flows in a region. Figure 1 is a graph of GDP (a), cumulative confirmed cases (b) of cities in Hubei, and GDP (c), cumulative confirmed cases (d) of cities outside Hubei. Figure 1 shows that there is a similar trend between GDP and the cumulative confirmed cases. Correlation analysis showed that the relationship between the two reached a very significant positive correlation. In general, cities with high GDP also have high confirmed cases. We divide the Chinese mainland cities into two categories: cities excluding Hubei Province and cities of Hubei Province, and describe them respectively. Cities outside Hubei Province, the correlation coefficient between GDP and confirmed cases is 0.7343 (reached 99% significance level). The cities with the top-20% GDP have a GDP of 60.65% and confirmed cases of 53.89%, meanwhile, the cities with the lowest-20% GDP have a GDP of 2.68% and confirmed cases of 3.30%. Cities of Hubei Province, as the origin place of the disease, their GDP and confirmed cases also have a more significant positive correlation, with a correlation coefficient of 0.9495 (reached 99% significance level).

**Demonstration from China**

Figure 1 shows that there is a significant positive correlation between GDP and confirmed cases in different regions of an epidemic area. According to this finding, the epidemic prevention and control in different economic development areas in one region can be graded. In areas with high economic development level, the epidemic risk is high, and the more active and strict epidemic prevention measures should be taken. Such as: call on the public to reduce the flow as much as possible; do a well in personal protection. Generally speaking, areas with high economic development have relatively higher flow of people and logistics, which also is a way of transmission for viruses. Therefore, during the epidemic period, we should minimize the flow, reduce the aggregation and prevent the virus from spreading among people. In particular, personal protection is very important to prevent infection, according to China's experience. However, some agencies recommend that healthy people do not wear face masks. But in fact, the role of face masks is mainly to block the virus, not treat the disease. Wearing a mask can prevent outside viruses from entering healthy people. It can also be seen from China's practical experience that wearing face masks has a significant effect on reducing the spread of the virus among people.

In developing areas, the risk of epidemic is lower than that of high economic development areas, but low risk is not without risk.

It is necessary to strengthen the publicity of the people and improve the awareness of prevention. Establish the concept of "if you don't pay attention to viruses, they will pay attention to you", guide the public to do a well personal protection, and isolate the cases early to prevent the epidemic from spreading internally.

https://www.who.int/news-room/detail/07-03-2020-who-statement-on-cases-of-covid-19-surpassing-100-000

**Figure 1** Cartogram of prefectural city cumulative confirmed cases during two incubations since Wuhan was sealed. (Note: Area is proportional to GDP in 2018 at the prefectural level.)

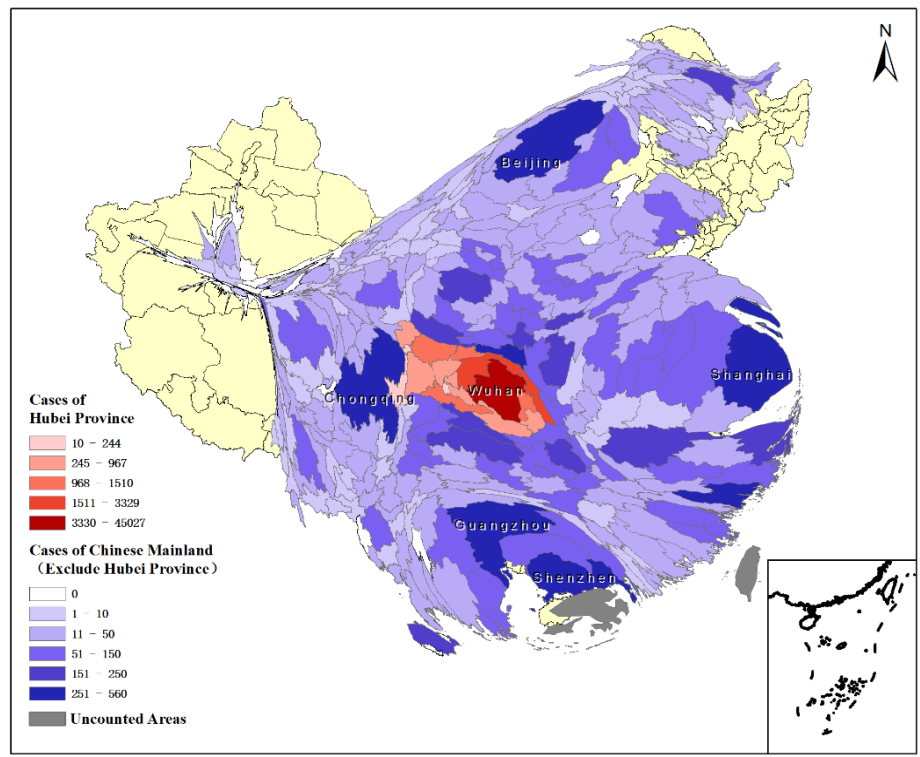